\documentclass[preprint,amsmath,amssymb]{revtex4}
\usepackage{epsfig}

\begin{document}

\title{Nanoelectromechanical systems}

\author{M. P. Blencowe}
\affiliation{Department of Physics and Astronomy, 6127 Wilder 
Laboratory,\\
Dartmouth College, Hanover, NH 03755, USA}

\date{\today}

\begin{abstract}

    Nanoelectromechanical systems (NEMS) are nano-to-micrometer scale 
    mechanical resonators coupled to electronic devices of similar 
    dimensions. NEMS show promise for fast, ultrasensitive force 
    microscopy and for deepening our understanding of how 
    classical dynamics arises by approximation to quantum dynamics. 
    This article begins with a survey of NEMS and then
    describes certain aspects of their classical dynamics. 
    In particular, we show that for weak coupling 
    the action of the electronic device on the mechanical resonator can 
    be effectively that of a thermal bath, this despite the device being a 
    driven, far-from-equilibrium system. 
    
\end{abstract}

\maketitle

\section{Introduction} \label{intro}

A nanoelectromechanical system (NEMS)  consists of a 
nanometer-to-micrometer (micron) scale mechanical resonator that is 
coupled to an electronic device of comparable 
dimensions~\cite{roukes1,roukes2,roukes3,cleland1,blick1,blencowe1}. The 
mechanical resonator may have a simple geometry, 
such as a cantilever (suspended beam clamped at one end) or a bridge 
(suspended beam clamped at both ends) and is fashioned out of 
materials such as silicon using similar lithographic techniques to 
those employed for fabricating integrated circuits. Because of their 
(sub)micron size, the mechanical resonators can vibrate at 
frequencies ranging from a few megahertz (MHz) up to around a 
gigahertz (GHz)~\cite{roukes4,geller1}; we are not normally accustomed to the 
idea of 
mechanical systems vibrating at such high radio-to-microwave 
frequencies. 

The coupling to the electronic device  can be achieved electrostatically by 
applying a voltage to a metal film deposited on the surface of the mechanical 
resonator.  One example of a coupled electronic device is a single electron 
transistor (SET) shown in figure~\ref{blencowe1}. 
Electrons quantum tunnel one at a time across the transistor from 
drain electrode to source electrode, 
driven by a drain-source voltage $V_{\rm ds}$ (we adopt the 
usual convention where positive charges flow  from source 
electrode to drain electrode and hence negative charges flow the 
opposite way). The magnitude of the resulting current depends 
on another voltage applied to a third, gate electrode, called the gate 
voltage $V_{\rm g}$. With the 
metallized mechanical resonator forming part of the gate electrode, 
motion of the former will modulate the gate voltage and hence the 
drain-source tunneling current, which is subsequently amplified and detected.

With the high frequencies and small inertial 
masses of the nanomechanical resonators, together with the 
ultrasensitive mechanical displacement detection capabilities of the coupled 
electronic devices, NEMS show great promise for metrology. One possible area of 
application is force microscopy, where a cantilever tip is 
scanned over a surface and the cantilever displacements measured as the tip 
interacts with the surface used to build up a force topography map. Of 
particular interest is the magnetic resonance force microscope (MRFM) which 
employs a ferromagnetic cantilever tip, enabling the mapping of 
unpaired electron and nuclear spin densities at and below the 
surface~\cite{sidles1}. Recently, single electron spin detection sensitivities 
were
achieved~\cite{rugar1,hammel1}; the potential applications of being able to 
determine chemical 
identity at the single molecule or atom level are numerous. And with 
the use of smaller, suitably-engineered NEMS MRFM devices, the 
higher mechanical frequencies might result in faster read-out times 
at equivalent or better sensitivities.

Another application is mass-sensing, where the mass of a small 
particle attaching itself to a nanomechanical resonator can be 
determined from the resulting vibrational frequency-shift of the resonator. 
Recently, attogram (`atto'$\equiv 10^{-18}$) detection sensitivities 
were achieved~\cite{ilic1,ekinci1}. With the use of suitably-engineered higher 
frequency 
NEMS, the detection of individual molecules may be possible at 
single-Dalton sensitivities (1 Dalton $=1.66\times 10^{-27}~{\rm 
kg}$, $1/12$ the mass of a ${\rm C}^{12}$ atom)~\cite{ekinci2}.

NEMS can be interesting in their own right as nontrivial dynamical 
systems. Because of the small inertial mass of the nanomechanical 
resonator and the strong electrostatic coupling to the closely integrated 
electronic device that can be achieved, individual electrons 
travelling through the latter can give significant displacement `kicks' to the 
mechanical resonator. In turn, the motion of the resonator will influence the 
electron current and so on. At cryogenic temperatures, certain electronic 
devices can behave in a quantum coherent fashion, existing in 
a quantum superposition of different charge states as electrons are 
transmitted through the device. Interacting with such a device, the 
mechanical resonator's centre-of-mass may be driven into a quantum 
state~\cite{blencowe1}, such as a superposition of separated position states. 
The
quantum nature of the coupled electromechanical system will manifest
itself in certain signatures of the measured current. Nanomechanical 
resonators comprise up to about ten billion atoms, so that by most 
standards such quantum effects would be deemed macroscopic. It is 
important to appreciate that we are here referring to quantum effects 
in `dirty real' devices that possess both many electronic and 
mechanical degrees of freedom, and which interact strongly with the 
surrounding environment consisting of photons, phonons, fluctuating 
(charge) defects in both the mechanical resonator and electronic device 
etc. The experimental and theoretical investigation of such systems 
will lead to a deeper understanding of how classical dynamics emerges 
as an approximation to quantum dynamics; NEMS straddle the microscopic 
quantum and macroscopic classical worlds.

In the first generation of experiments to probe 
the dynamics of NEMS 
(see, e.g., Refs.~\cite{knobel1,blencowe2,lahaye1,blencowe3,scheible1,
sazonova1,cleland2}), 
the mechanical resonator 
component has been found to behave 
classically, as might be expected; the experiments are not yet quite refined 
enough to observe quantum interference effects that are washed-out by the 
resonator's environment.  Despite this, the (semi)classical 
dynamics of NEMS has been found to be nontrivial and worthy of 
investigation. One line of investigation is to identify common features in the 
classical dynamics of the various NEM devices, so as to bring some 
degree of coherence to the field. Remarkably, 
under certain conditions of weak coupling and also wide separation of 
mechanical and electronic dynamical time-scales, the 
electronic device behaves effectively as a thermal bath: the mechanical resonator 
undergoes thermal brownian motion, characterized by a damping 
constant and effective temperature which are determined by the 
electronic parameters of the device~\cite{mozyrsky1,armour1}. The fact that the 
electronic 
device can be effectively replaced by a thermal bath is at first sight 
rather surprising, given that the voltage-driven electron current 
flowing through the device is a far-from-equilibrium, many 
electron state. The use of well-understood equilibrium concepts 
in the construction of theoretical models of less-well-understood non-equilibrium 
systems goes back to the early days of statistical mechanics and 
continues to find broad application.  For a recent review, see for 
example Ref.~\cite{casas1}.  

The outline of this article is as follows: section~\ref{survey} 
gives examples of various 
representative NEM devices that are being investigated. Section~\ref{SET} 
analyses the classical dynamics of the SET-mechanical resonator system, focusing 
on the effective equilibrium description in the weak-coupling regime.
Section~\ref{NEMS} 
describes the effective equilibrium dynamics of some of the other NEMS 
introduced in section~\ref{survey}.
Section~\ref{conclusion} gives concluding remarks.

\section{Survey of NEMS} \label{survey}

In this section we describe several representative NEMS. The 
explanations as to how they work will be qualitative and brief in nature, 
centering on schematic diagrams or electron microscope images of each 
device.

\begin{figure}[htbp]
\centering
\epsfig{file=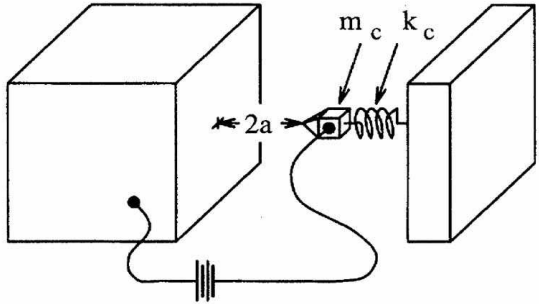,width=3.in} 
\caption {Scheme for a mechanically compliant tunneling 
electrode~\cite{schwabe1}. 
The  cantilever electrode is modeled as a harmonic oscillator with 
some spring constant $k_{c}$ and mass $m_{c}$. A voltage is applied 
across the gap $2a$ and the resulting current measured.}
\label{schwabe}
\end{figure}

Figure~\ref{schwabe} shows a schematic diagram of a mechanically compliant 
tunneling electrode taken from Ref.~\cite{schwabe1}. The tunneling 
electrode consists of a cantilever with metal tip which is placed 
close to the surface of a bulk metal counterelectrode. As is common in 
theoretical analyses of NEMS, the cantilever is simply modeled as a harmonic 
oscillator with some effective spring constant and mass. Electrons tunneling 
across the gap between the metal tip and surface cause the cantilever 
to recoil, while in turn the cantilever's motion affects the electron 
tunneling probability and hence the measured tunneling current. Thus, 
we have a coupled electromechanical system which should display some 
interesting dynamical properties, depending on the applied voltage 
across the gap, gap size, cantilever's mass, 
spring constant etc. There have been several theoretical 
investigations of the mechanical tunneling electrode 
and related schemes
~\cite{bocko1,yurke1,presilla1,onofrio1,schwabe1,mozyrsky1,clerk1,smirnov1,xue1}.  
In a sense, the scanning tunneling microscope~\cite{binnig1} is an experimental 
realisation, 
since a tunneling electrode cannot be made absolutely rigid. However, a 
device with a deliberately compliant, low-mass tunneling electrode, such that 
the tunneling electrons themselves cause significant recoil of the electrode, 
has yet to be demonstrated. 

\begin{figure}[htbp]
\centering
\epsfig{file=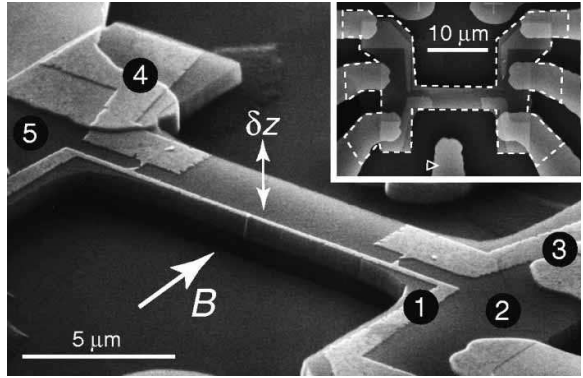,width=3.in} 
\caption {SEM micrograph of QPC displacement detector, showing the 
suspended beam with electrodes on the surface~\cite{cleland3}. 
The gate electrodes are labeled `1' and `3'. The 
point contacts forming the narrow constriction are located just above 
label `1'. The drain and source electrodes are labeled `5' and `2'. 
The indicated magnetic field is used to actuate the beam; the Lorentz 
force on the electron current causes flexing displacements in the 
indicated $z$ direction.}
\label{cleland}
\end{figure}

Figure~\ref{cleland} shows a scanning electron microscope (SEM) 
micrograph of a quantum point contact (QPC) displacement detector 
developed by Andrew Cleland's group at UC Santa Barbara~\cite{cleland3}.
The detector comprises a suspended beam etched from a single-crystal
gallium arsenide (GaAs) heterostructure with a fundamental resonance 
frequency for out-of-plane flexural motion of about $1.5~{\rm MHz}$. 
Within the beam is a thin layer of free electrons, called a 
two-dimensional electron gas, (see, e.g., Ref~\cite{challis1} for 
an elementary review of 2DEG and other low-dimensional semiconductor systems) 
which 
forms a current when a drain-source voltage is applied across the ends 
of the beam. Applying also a sufficiently negative voltage to the 
metal gate electrodes on the surface of the beam  expels the electrons 
from 
directly beneath the elecrodes so that they can only flow through the 
narrow, electrostatically defined constriction between the point 
contacts. The constriction can be made narrow enough such that it is 
comparable to the electrons' Fermi wavelength; hence the name 
`quantum point contact'. Because GaAs is a piezoelectric material, mechanical 
strain in the flexing beam will induce a polarization electric field in the beam.
This induced polarization field has the same 
effect as applying a gate voltage, hence modulating the current 
flowing through the QPC. In turn, the fluctuating electric field due 
to the flowing current will induce a mechanical strain. Thus, the QPC 
displacement detector is another example of a coupled 
electromechanical system.      

\begin{figure}[htbp]
\centering
\epsfig{file=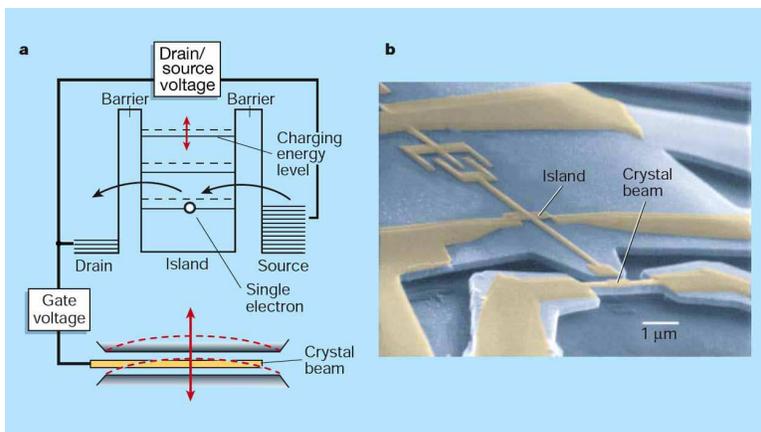,width=4.in} 
\caption {(a) Cartoon illustrating the operation of the SET 
displacement detector. The indicated charging energy levels represent 
the energy cost arising from the change in stored electric field 
energy as one or more electrons tunnel onto the island; the 
discreteness of the levels are not a quantum effect, but rather the incremental 
energy cost to put increasing numbers of electrons on the island at 
the same time. (b) False-coloured SEM micrograph showing the 
suspended, doubly-clamped beam and SET~\cite{knobel1}. 
The substrate and beam are 
fashioned from GaAs (blue regions), and the SET and beam gate electrodes are thin 
layers of aluminum (yellow regions), 
with aluminum oxide forming the tunnel barriers. 
The beam is located $0.25$~microns away from the island electrode. The 
measured fundamental flexural frequency for in-plane motion is about 
$116~{\rm MHz}$.}
\label{blencowe1}
\end{figure}

\begin{figure}[htbp]
\centering
\epsfig{file=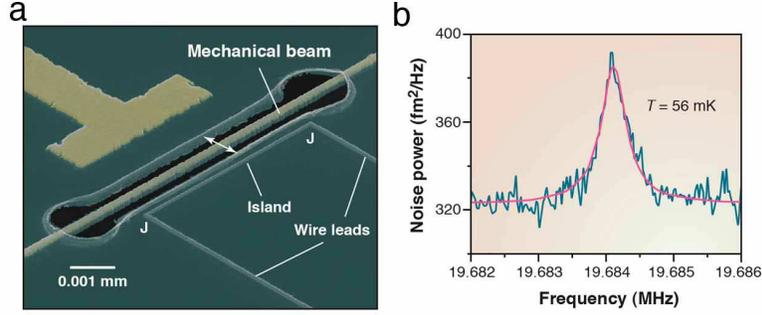,width=4.in} 
\caption {(a) SEM micrograph showing the mechanical beam and SET wire 
drain-source wire leads and island~\cite{lahaye1}. 
Electrons tunnel one at a time 
across the electrically-insulating junctions located at the corners 
(J). The beam and surrounding substrate are fashioned from a 
silicon-nitride (SiN) membrane, a commonly-used material for NEMS 
because of its high strength-to-mass density ratio. The beam is coated 
with a layer of gold, forming the gate electrode, while the SET island 
and leads are made from aluminum. The  mechanical resonator is 
located $0.6$~microns away from the island. The large, fixed gate 
electrode at the upper left is used for actuation.  (b) SET-detected 
noise power spectrum converted to displacements-squared. The 
Lorentzian peak due to the mechanical beam's thermal brownian motion 
is clearly seen above the amplifier noise background.  The measured 
fundamental frequency for in-plane motion is about 19.7~MHz.}
\label{blencowe2}
\end{figure}

\begin{figure}[htbp]
\centering
\epsfig{file=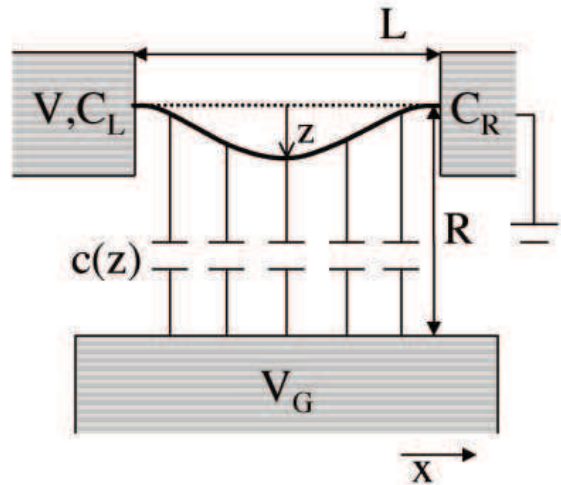,width=3.in} 
\caption {Scheme for a SET with mechanically-compliant island formed 
out of a suspended carbon nanotube~\cite{sapmaz1}. As a result of the 
electrostatic 
coupling between the voltage-biased gate and island, the position of the latter 
fluctuates as electrons tunnel on and off.}
\label{sapmaz}
\end{figure}

Figures~\ref{blencowe1},~\ref{blencowe2}, and \ref{sapmaz} show 
various realisations of the single electron transistor (SET) 
displacement detector. The device in figure~\ref{blencowe1} was 
developed by Cleland's group~\cite{{knobel1},{blencowe2}}, while the 
device in figure~\ref{blencowe2} was developed by Keith Schwab's 
group at the Laboratory for Physical Sciences, U 
Maryland~\cite{lahaye1,blencowe2}. 
Figure~\ref{sapmaz} shows a scheme being developed by Herre 
van der Zant's group at Delft~\cite{sapmaz1}. The latter device 
differs from the first two in that the SET island is 
mechanically-compliant instead of the gate electrode. A 
closely-related transistor device integrating a suspended carbon nanotube was 
recently demonstrated by Paul McEuen's 
group at Cornell~\cite{sazonova1}.   The basic operating 
principle of the SET displacement detector is illustrated in the 
cartoon (figure~\ref{blencowe1}a). The source and drain aluminum electrodes 
are electrically insulated from the island electrode by an oxide 
layer. However, because the oxide layer is very thin, electrons can quantum 
tunnel across the oxide barriers, from drain-to-island-to-source, 
constituting a current which is measured after subsequent 
amplification stages. In order for an electron to be able to tunnel 
from, e.g., the drain to the island electrode, the total work 
done by the drain-source and gate voltages 
must exceed the accompanying 
change in the stored electric field energy due to the redistribution 
of the charges on the various electrodes (called the 
`charging energy'); the difference between 
the work done and the charging energy  gives the net energy 
gained by the tunneling electron, which must be positive. 
Because of the small sizes of the 
electrodes, their capacitances are small and so the charging energy 
is large. Hence, a large enough drain-source voltage must be 
applied in order to meet this charging energy cost so that electrons 
can tunnel through the device. If, however, the voltage is not large enough 
to overcome the charging energy required to put more than one electron
simultaneously on the island, then only one electron can tunnel on and off 
the island at a time (hence the name `single electron transistor'). 

The charging energy can be offset by the applied gate voltage, so that 
varying the gate voltage modulates the tunneling current. Alternatively, 
by applying a fixed voltage to the mechanically-compliant beam gate electrode, 
motion of the latter will also modulate the tunneling current.  
Figure~\ref{blencowe2}b shows an example from Schwab's 
group of the SET-detected signal of the mechanical beam undergoing thermal 
brownian motion. The area under the peak after subtracting off the  
white-noise background (which mostly originates from the subsequent 
amplification stages) gives a measured mean-squared displacement $\langle 
x^2\rangle$ of about $2\times 10^{-13}~{\rm m}$, with a position 
detection sensitivity set by the white-noise floor of about $10^{-13}~{\rm m}$.
To put these numbers into perspective, this SET is able to detect 
displacements as small as one-thousandth the diameter of a hydrogen 
atom. Furthermore, such sensitivities are within an order of magnitude 
of the quantum zero-point displacement uncertainty of the mechanical beam. 

Using the classical equipartition of mechanical energy for a harmonic 
oscillator, $\langle 
E\rangle=m\omega^{2}\langle x^{2}\rangle =k_{\rm B} T$, it is 
possible to directly determine the temperature of the mechanical beam 
from the measured mean-squared displacement, resonant frequency 
and effective motional mass (which can be estimated from the beam 
dimensions and mass density). The data in figure~\ref{blencowe2}b 
corresponds to a beam temperature of about 60~mK. This value sets a 
record for the lowest directly measured temperature of a 
nanomechanical resonator. The actual, base temperature  of the 
refrigerator is about 35~mK, suggesting that there is some local 
heating of the beam. Most likely, the energy released by the 
tunneling electrons in the SET is not dissipating quickly enough.

In addition to heating the mechanical beam resonator,  
the tunneling electrons will directly influence the motion of 
the resonator: 
the fluctuating island voltage  due to electrons tunneling 
on and off the island will produce a back reaction force on the 
electrostatically coupled resonator. The magnitude of this force noise 
increases with applied gate voltage. The displacement detection 
sensitivity also increases, so that there is an optimum gate voltage 
bias point which is neither too large nor too small. On the other 
hand, if our interest is not in displacement detection, but rather 
the investigation of the coupled dynamics of NEMS, then the gate 
voltage should be increased beyond this optimum bias point so that 
there is a large back reaction on the resonator . Motion of the 
resonator then modulates the tunneling current and in turn the 
fluctuating current drives the mechanical resonator. Experiments are 
underway in Schwab's group to explore the coupled dynamics in this 
back reaction-dominated regime.     

\begin{figure}[htbp]
\centering
\epsfig{file=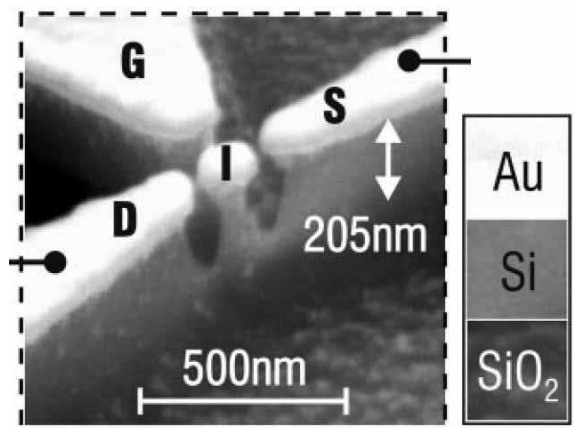,width=3.in} 
\caption {SEM micrograph of the silicon nanopillar with gold island 
(I) and source (S) and drain (D) electrodes~\cite{scheible1}. 
The gate (G) electrode 
was not used in the experiment. The current is detected 
from the drain electrode after amplification. The nanopillar has a 
fundamental flexural frequency of 367~MHz.}
\label{scheible}
\end{figure}

Figure~\ref{scheible} shows a nanomechanical charge shuttle developed 
by Dominik Scheible at Ludwig Maximilians University, Munich, and 
Robert Blick at the University of Wisconsin~\cite{scheible1}. The 
mechanical shuttle element is a nanopillar fashioned out of silicon 
with a conducting island at the top made out of gold. An ac voltage 
is applied to the source electrode at a frequency close to 
the fundamental flexural frequency of the pillar. When there is an 
excess charge on the island, the ac voltage exerts a force on the 
pillar, driving it into mechanical oscillation at the ac frequency. If 
the ac voltage amplitude is sufficiently large, then the shuttle will 
deflect close enough to the source and drain electrodes such that 
electrons can tunnel between the island and the electrodes. The 
pillar then shuttles charge between the two electrodes, producing a 
current that is detected at the drain electrode. The number 
of electrons that tunnel and the tunnel direction depend on the 
magnitude and sign of the drive voltage at the instant of closest 
approach to a given electrode, i.e., on the phase lag between the 
oscillating mechanical motion and drive voltage. The phase lag in 
turn depends on the drive frequency relative to the  
fundamental frequency of the shuttle. Thus, the magnitude and 
direction of the shuttle current can be controlled by varying the 
drive frequency. The charge shuttle bears some resemblance to the SET 
with mechanically compliant island shown in figure~\ref{sapmaz}. 
However, as the name 
implies, in contrast to the above-described NEM devices
mechanical motion of the shuttle is essential for an electron current 
to flow; the electronic and mechanical degrees of freedom are 
inextricably linked in their dynamics. Ref.~\cite{shekhter1} gives a
review of charge shuttle systems.

\begin{figure}[htbp]
\centering
\epsfig{file=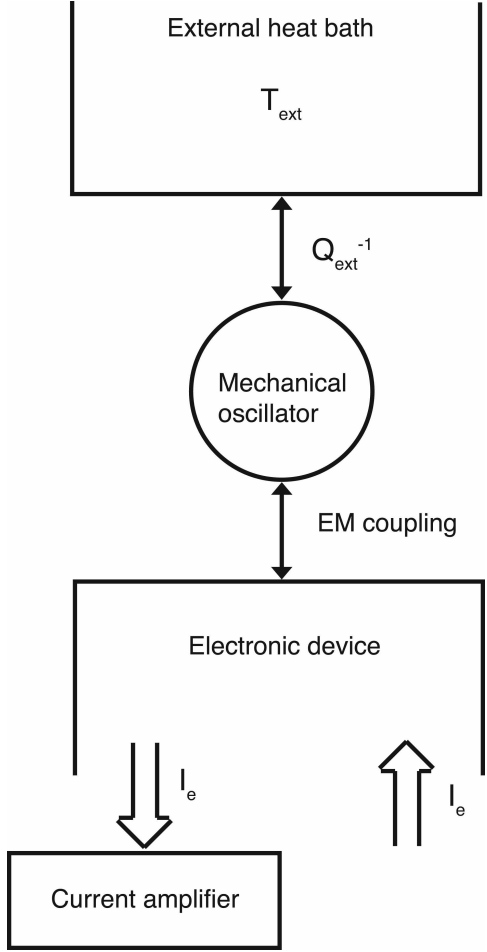,width=2.5in} 
\caption {Scheme of a generic NEMS.}
\label{genericnems}
\end{figure}

In this section, we  have introduced several examples of NEM devices. The 
following sections will explore certain aspects of their coupled 
dynamics, with a central goal being to identify commonalities in their 
dynamical behaviour. In this respect, it will be useful to view a NEM 
device according to the generic scheme of figure~\ref{genericnems}. 
The mechanical resonator, which will be simply modeled in its lowest 
fundamental vibrational mode as a harmonic oscillator, is an 
open system that is coupled to two energy reservoirs. The electronic device  
through which an electron current flows constitutes one of the 
reservoirs where the 
current exchanges energy with the  oscillator via the 
electromagnetic interaction.  All degrees of freedom apart from those 
of the electronic device and the oscillator constitute the second, 
`external' reservoir. These degrees of freedom consist, for example, 
of higher vibrational modes, fluctuating defects etc. within the 
mechanical resonator and air molecules, photons etc. impinging the 
surface of the resonator. These external degrees of freedom are 
simply modeled as one infinitely large thermal equilibrium
bath at some temperature 
$T_{\rm ext}$. If the oscillator is initially in an excited state, then 
in the absence of the electronic device it will 
lose energy to this bath at a rate which can be expressed as 
$\omega/Q_{\rm ext}$, where $\omega$ is the oscillator frequency and 
$Q_{\rm ext}$ is the quality factor. Nanomechanical resonators are 
typically found to have quality factors between 
$10^{3}-10^{5}$~\cite{roukes3}, so that a resonator oscillates at 
least a few thousand cycles before its amplitude has substantially 
decayed from the initial excited amplitude.

We shall pay particular attention to the dynamics of the oscillator 
system as a result of its interactions with these two
reservoirs. In this respect, we shall envisage a separate, idealized 
direct probe of 
the oscillator dynamics allowing perfect measurement of its position 
and velocity coordinates without affecting its dynamics. 
In actual NEMS experiments, 
information about the mechanical resonator dynamics is obtained by 
measuring the electron current: the electronic device is after all 
usually intended as a displacement detector. For some theoretical 
investigations about how the dynamics of the mechanical resonator 
manifests itself in the current signal output, see, e.g., 
Refs.~\cite{clerk1,armour2,flindt1,clerk2}.

\section{The SET-nanomechanical resonator system} \label{SET}

In this section we describe the coupled classical dynamics of the 
SET-mechanical resonator system. The coupled dynamics of other NEMS 
will be considered in section~\ref{NEMS}. Referring to the generic 
scheme of figure~\ref{genericnems}, we shall in the first instance 
omit the coupling to the external thermal bath (i.e., $Q_{\rm 
ext}\rightarrow\infty$); we are interested in elucidating the `pure', 
coupled dynamics of the  oscillator interacting with the SET only. 
For example, is this 
system stable in the sense that the tunneling electron current removes 
energy from an initially excited oscillator causing damping, or do the tunneling 
electrons dump energy into the oscillator driving it to progressively larger 
amplitudes? The answer is not obvious a priori.   

\subsection{The master equation} \label{master}

We seek some form of manageable equation  of motion which describes the 
SET-mechanical resonator system. The logical starting point is the time-dependent 
Schr\"{o}dinger equation with Hamiltonian involving the relevant 
microscopic degrees of freedom, in particular the conduction electrons 
in the various SET electrodes, and the fundamental 
oscillator mode (recall we are neglecting for the time being the external 
environment of the oscillator mode). One then proceeds through 
various stages of approximation to arrive at simpler-to-handle
classical statistical equations  of motion along with some conditions for 
their validity. However, for NEMS at least, this 
procedure still needs to be properly developed and 
is essential to the future goal to achieve a 
deeper understanding of how classical dynamics emerges from quantum dynamics 
by approximation in these systems.

In the absence of such a properly-defined procedure, we shall instead 
use physical intuition to guide us directly towards writing down the classical 
statistical equations of motion. There is a long tradition employing such an 
approach going back to Boltzmann and his famous equation. One advantage 
is that such equations often accurately capture the statistical dynamics as 
verified by experiment. Another particular advantage in our case
is that the dynamics in 
the  regime of strong electromechanical coupling is easier to 
analyze than in the full quantum theory. The general disadvantage is that 
the precise conditions for the validity of the classical equations of motion 
are not available, with the consequence that important terms can 
occasionally be overlooked. This is believed not to be a problem for the 
SET-oscillator classical equations  which we shall shortly write down. 

\begin{figure}[htbp]
\centering
\epsfig{file=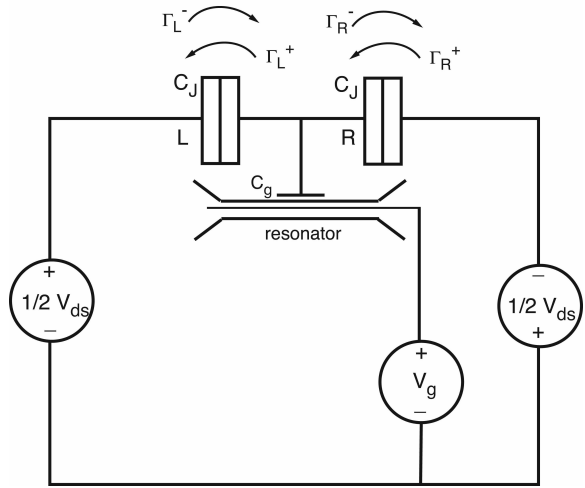,width=3.in} 
\caption {Circuit diagram of the SET-mechanical resonator system. The 
capacitances of the left ($L$) and right ($R$) tunnel junctions are 
assumed identical, denoted as $C_{J}$. The gate capacitance
formed by the adjacent resonator and island electrodes is denoted 
$C_{g}$. As the resonator flexes,  the distance between the two electrodes 
varies, hence changing $C_{g}$. With the resonator electrode kept 
fixed at the gate voltage $V_{g}$, the changing $C_{g}$ gives rise to 
a varying island potential, hence modulating the tunneling current. 
For the indicated drain-source voltage ($V_{\rm ds}$) polarity, the usual 
direction for electron flow is from right to left, governed by the 
tunnel rates $\Gamma_{R}^{+}$ and $\Gamma_{L}^{+}$.}
\label{SETcircuit}
\end{figure}

What minimal set of coordinates is required to describe the 
mechanical resonator interacting with the SET? Modeled as a single 
oscillator mode, the resonator's state is described by a position $x$ 
and velocity $v$ coordinate. With respect to the oscillator, the 
relevant SET coordinate is the total number $N$ of excess electrons on 
its island; depending on $N$, the electrostatic force will pull the 
 oscillator towards the island by different amounts. (If we 
were to also describe the SET drain-source current, we would 
need an additional counter coordinate which keeps track of the 
number of electrons entering (or leaving) the SET).  
Because tunneling is a random process, the coordinates $N$, $x$ and 
$v$ will fluctuate. Thus, the equations of motion can either take the 
form of stochastic differential equations involving the random, 
time-varying functions  $N(t)$, $x(t)$ and $v(t)$ (see e.g., 
Ref.~\cite{lemons1} for an introduction on how to describe stochastic 
processes), or we can write down a deterministic equation involving 
the probability density function $P_{N}(x,v,t)$. In the latter 
formulation, $P_{N}(x,v,t)dxdv$ is interpreted as the probability 
at time $t$ of picking out from a large number of identical SET-oscillator 
systems (an ensemble), one system with island number $N$ and with position and 
velocity coordinates in the respective intervals $[x,x+dx]$ and $[v,v+dv]$.
We shall choose to express the dynamics in terms of the probability 
density function, rather than in terms of stochastic coordinates. 
Referring to the circuit diagram for the coupled SET-resonator 
system (figure~\ref{SETcircuit}), we have~\cite{armour1}
\begin{eqnarray}
    \frac{dP_{N}}{dt}&=&\{ 
    H_{N},P_{N}\}-\left(\Gamma^{-}_{L}+\Gamma^{+}_{R}\right) P_{N}+
    \left(\Gamma^{+}_{L}+\Gamma^{-}_{R}\right) P_{N+1}
    \label{mastereq1}\\
    \frac{dP_{N+1}}{dt}&=&\{ 
    H_{N+1},P_{N+1}\}-\left(\Gamma^{+}_{L}+\Gamma^{-}_{R}\right) 
    P_{N+1}+
    \left(\Gamma^{-}_{L}+\Gamma^{+}_{R}\right) P_{N},
    \label{mastereq2}
\end{eqnarray}
where $H_{N(N+1)}$ is the oscillator Hamiltonian for the resonator in the 
background electrostatic potential of the SET 
with $N(N+1)$ electrons on the island, $\{{\cdot},{\cdot}\}$ is the 
Poisson bracket and $\Gamma^{\pm}_{L(R)}$ are the electron tunneling 
rates to the left ($+$) or to the right ($-$) across the left ($L$) 
or right ($R$) tunnel junctions. Here, we are assuming that the 
voltages are chosen such that the number of excess electrons on the 
island fluctuates between $N$ and $N+1$ only. 

Equations~(\ref{mastereq1}) and (\ref{mastereq2}) describing the 
evolution of the probability distribution $P_{N(N+1)}(x,v,t)$ are
commonly called  `master equations'. 
They are coupled, first-order partial differential 
equations in the variables $x$, $v$ and $t$. Once an initial 
probability distribution has been specified, then these equations 
uniquely determine  the subsequently evolving probability distribution. 
For example, the distribution at some initial time $t_{0}$ might take the form
$P_{N}(x,v,t_{0})=\delta (x) \delta (v)$ and $P_{N+1}(x,v,t_{0})=0$, 
meaning that at time $t_{0}$ all SET-oscillator systems within the 
ensemble are in the same state, with the oscillator's position and 
velocity being $x=0$ and $v=0$, respectively, and the SET island 
number being $N$. If the tunneling rate terms were absent in 
equations~(\ref{mastereq1}) and (\ref{mastereq2}) then the evolving probability 
distribution 
would remain a delta function, describing deterministic harmonic 
oscillator motion with the island number remaining fixed at $N$. 
Choosing the origin of the coordinate $x$ 
to coincide with the equilibrium position of the oscillator
when there are $N$ electrons on the island, the oscillator 
Hamiltonians take the form
\begin{eqnarray}
    H_{N}&=&\frac{p^{2}}{2m}+\frac{1}{2} m\omega^{2}x^{2}
    \label{oschamiltonian1}\\
    H_{N+1}&=&\frac{p^{2}}{2m}+\frac{1}{2} m\omega^{2}(x-x_{0})^{2},
    \label{oschamiltonian2}
\end{eqnarray}
where $x_{0}$ is the distance between equilibrium positions of the 
oscillator with $N$ and $N+1$ electrons on the island. 
With the tunneling rate terms present, however, the evolving probability 
distribution begins to spread. The fluctuating electron island number 
due to electrons tunneling on and off the island causes the 
deterministic evolution of the oscillator to be interrupted at random 
times by a sudden shift $\pm x_{0}$ in the origin of the harmonic 
potential experienced by the oscillator.

The placings and the signs in front  of the various tunneling rate 
terms in the master equations~(\ref{mastereq1}) and (\ref{mastereq2}) are easily
understood from figure~\ref{SETcircuit}. For example, 
in equation~(\ref{mastereq1}), the first two rates $\Gamma^{-}_{L}$ 
and $\Gamma^{+}_{R}$ correspond to tunneling onto the island and so 
decrease the likelihood that the island number remains at $N$; hence 
the `minus' sign. On the other hand, the second two rates $\Gamma^{+}_{L}$ 
and $\Gamma^{-}_{R}$ correspond to tunneling off the island and so 
increase the likelihood that the island number becomes $N$; hence the 
`plus' sign. The tunneling rates can be derived using Fermi's Golden 
Rule~\cite{amman1} and take the form
\begin{equation}
\Gamma^{\pm}_{L(R)}=\frac{1}{e^{2}R_{J}}
\frac{E^{\pm}_{L(R)}}{1-e^{-E^{\pm}_{L(R)}/k_{\rm B}T_{e}}},
\label{tunnelrates1}
\end{equation}
where $R_{J}$ is the effective tunnel junction resistance (assumed 
the same for each junction), $T_{e}$ is the temperature of the 
source, drain and island electron reservoirs (assumed the same) and 
$E^{\pm}_{L(R)}$  is the energy gained by a 
single electron tunneling to the left ($+$) or to the right ($-$) across the 
left ($L$) or right ($R$) junction. When the electron temperature 
$T_{e}$ is small compared to the single electron charging energy, i.e, $k_{\rm B} 
T_{e}\ll e^{2}/2C_{\Sigma}$ (where total SET capacitance is $C_{\Sigma}=2C_{J}+C_{g}$), 
then equation~(\ref{tunnelrates1}) becomes approximately
\begin{equation}
\Gamma^{\pm}_{L(R)}=
\frac{1}{e^{2}R_{J}}E^{\pm}_{L(R)}\Theta(E^{\pm}_{L(R)}),
\label{tunnelrates2}
\end{equation}
where $\Theta(\cdot)$ is the Heaviside step function. Thus, a given 
tunnel rate is only non-negligible provided the associated $E$ is 
positive. For micron-scale SETs, $C_{\Sigma}\sim 1~{\rm fF}~(\equiv 
10^{-15}~{\rm F}$), giving $T_{e}\ll 1~{\rm K}$. In most experiments, 
this condition is satisfied and so from now on  we will use 
expression~(\ref{tunnelrates2}) for the rates. 
For the indicated drain-source voltage polarity in 
figure~\ref{SETcircuit}, rates $\Gamma_{R}^{+}$ and $\Gamma_{L}^{+}$ 
describing tunneling to the left are non-negligible, while the rates
$\Gamma_{R}^{-}$ and $\Gamma_{L}^{-}$ for tunneling to the right are 
exponentially suppressed. Note, however, that through the $E$'s, 
the tunnel rates also 
depend on the oscillator's position $x$; it is possible that, for 
large enough amplitude displacements and applied gate voltages, 
a given $E$ changes sign hence switching 
on or off the corresponding tunnel process and changing the tunneling 
direction.

The above master equation pair (\ref{mastereq1}) and (\ref{mastereq2})
is clearly classical. The probability 
distribution characterizes the classical statistical uncertainty in 
ones knowledge of the precise coordinates $x$, $v$ of the oscillator 
and island electron number. The equations do not entertain the 
possibility of quantum superpositions between different island number 
states, different oscillator position/velocity states, or entangled 
oscillator-island states. Quantum mechanics only enters in the 
determination of the effective tunnel junction resistances $R_{J}$
appearing in  the rate expressions~(\ref{tunnelrates1}), essentially providing 
the randomness of the transition rates. Refering to the discussion at 
the beginning of this section, the classical master equation  can in principle 
be derived from the Schr\"{o}dinger equation describing the 
time-evolution of the density matrix characterising the quantum state of the 
harmonic oscillator and conduction electrons in the SET electrodes. 
One step in the derivation is to `trace-out' the many electron state, 
leaving in the SET state description just the total island number. 
This step introduces irreversibility into the equations, as manifested 
by the tunneling rate terms. The two key 
approximation steps out of which a classical description emerges are
(1) assume a large enough SET island such that electron energy level 
quantization can be neglected; (2) assume large enough tunnel junction 
resistances such that the electron tunnels incoherently between the 
drain/source and island electrodes. 
Another approximation step that 
is made is to assume that the timescale for polarization charges 
on the electrode surfaces to re-equilibrate in response to a tunneling 
event is negligible as compared with the characteristic time between 
tunnel events given by $\tau_{\rm tunnel}=e R_{J}/V_{\rm ds}$. 
This is reflected in the fact that the rate of change of the 
probability densities at time $t$ in the master equation 
are governed by transition rates that are weighted by 
probability densities at the same time $t$ and not at earlier times. 
This is called the `Markovian' approximation. With the 
oscillator excluded, the above 
master equation arising from the various just-described approximations  
often goes by the name of the SET `orthodox model' (see, e.g., chapter 
4 of Ref.~\cite{ferry1} for an accessible discussion).          

\subsection{The steady state solution} \label{steady}

In solving the master equations (\ref{mastereq1}) and 
(\ref{mastereq2}), we will first determine the steady state behaviour 
for $t\rightarrow\infty$. Assuming sufficiently large $V_{\rm ds}$ 
and sufficiently weak coupling between the oscillator and SET such 
that the energy terms $E_{R}^{+}$ and $E_{L}^{+}$ are always positive, 
then we need only consider the rates $\Gamma_{R}^{+}$ and $\Gamma_{L}^{+}$ as 
defined in equations~(\ref{tunnelrates2}) with the step function 
omitted. It is convenient to express the master equations in terms of 
dimensionless coordinates, since in dimensionless form the essential 
parameters governing the dynamics are more clearly expressed. 
Expressing the time coordinate in tunneling time units $\tau_{\rm tunnel}$,  
the position coordinate in shift 
units $x_{0}$ and the velocity  coordinate in units $x_{0}/\tau_{\rm tunnel}$, 
the master equations take the form
\begin{eqnarray}
    \dot{P}_{N}&=&\epsilon^{2} x\frac{\partial P_{N}}{\partial v}- 
    v\frac{\partial P_{N}}{\partial 
    x}+\tilde{E}_{L}^{+}P_{N+1}-\tilde{E}_{R}^{+}P_{N} 
    \label{dimensionlessmastereq1}\\
    \dot{P}_{N+1}&=&\epsilon^{2} (x-1)\frac{\partial P_{N+1}}{\partial v}- 
    v\frac{\partial P_{N+1}}{\partial 
    x}-\tilde{E}_{L}^{+}P_{N+1}+\tilde{E}_{R}^{+}P_{N}
    \label{dimensionlessmastereq2},
\end{eqnarray}
where the dimensionless parameter 
$\epsilon=\omega \tau_{\rm tunnel}$ characterizes the separation 
between the  oscillator and SET dynamics timescales.
The dimensionless energy terms are obtained by dividing 
$E_{L(R)}^{\pm}$ by $eV_{\rm ds}$:
\begin{eqnarray}
    \tilde{E}_{L}^{+}&=& -\frac{e}{C_{\Sigma}V_{\rm ds}}(N_{g}-N-1/2)-\kappa N 
    +1/2 -\kappa x\label{dimensionlessenergy1}\\
    \tilde{E}_{R}^{+}&=& +\frac{e}{C_{\Sigma}V_{\rm ds}}(N_{g}-N-1/2)+\kappa N 
    +1/2+\kappa x,\label{dimensionlessenergy2}
\end{eqnarray}
where $N_{g}=C_{g}V_{g}/e$ is the polarization charge induced by the 
gate voltage and the dimensionless parameter 
$\kappa=m\omega^{2}x_{0}^{2}/(e V_{\rm ds})$ 
characterizes the coupling strength between the oscillator and the 
SET. The shift coordinate has the explicit form 
$x_{0}=-eN_{g}/(C_{\Sigma} m\omega^{2}d)$, where $d$ is resonator-island 
electrode gap, so that the coupling strength can be controlled by varying the 
gate voltage $V_{g}$; in particular, $\kappa$ depends quadratically 
on $V_{g}$. As we shall see, parameters $\epsilon$ and $\kappa$ are 
key to describing the coupled SET-oscillator dynamics.

From (\ref{dimensionlessmastereq1}) and 
(\ref{dimensionlessmastereq2}), we can derive equations for the 
various moments $\langle x^{n}v^{m}\rangle_{N(N+1)}$ of the 
probability distribution, where
\[
\langle x^{n}v^{m}\rangle_{N(N+1)}=\int dx \int dv\ 
    x^{n}v^{m}P_{N(N+1)}(x,v,t).
\]
Solving for the moments is a more manageable task than trying to solve 
for the full probability distribution all at once. The equations for 
the moments are
\begin{eqnarray}
    \langle \dot{x^{n}v^{m}}\rangle_{N}&=&-m\epsilon^{2}
    \langle x^{n+1}v^{m-1}\rangle_{N}+n\langle 
    x^{n-1}v^{m+1}\rangle_{N} + E_{L}\langle x^{n}v^{m}\rangle_{N+1}\cr
    &&-E_{R}\langle x^{n}v^{m}\rangle_{N}
    -\kappa\left(\langle x^{n+1}v^{m}\rangle_{N}+\langle 
    x^{n+1}v^{m}\rangle_{N+1}\right)\label{momenteq1}\\
    \langle \dot{x^{n}v^{m}}\rangle_{N+1}&=&-m\epsilon^{2}
    \left(\langle x^{n+1}v^{m-1}\rangle_{N+1}-\langle 
    x^{n}v^{m-1}\rangle_{N+1}\right)+n\langle x^{n-1}v^{m+1}\rangle_{N+1}\cr
    &&-E_{L}\langle x^{n}v^{m}\rangle_{N+1}+E_{R}\langle 
    x^{n}v^{m}\rangle_{N}\cr
    &&+\kappa\left(\langle x^{n+1}v^{m}\rangle_{N}+\langle 
    x^{n+1}v^{m}\rangle_{N+1}\right),
    \label{momenteq2}
\end{eqnarray}
where the $\kappa$-dependent coupling terms have been pulled outside the 
energy $E$ terms and the  `$\sim$' and `+' symbols on the latter have 
been dropped for notational convenience. If the SET damps the 
oscillator, then in the limit $t\rightarrow\infty$ the various moments 
approach constant values. Thus, we seek a possible solution to 
(\ref{momenteq1}) and (\ref{momenteq2}) with 
$\langle \dot{x^{n}v^{m}}\rangle_{N(N+1)}=0$. Fortunately, these equations 
can be solved exactly for the time-independent moments. For $n+m=0,1$, 
we find~\cite{armour1}
\begin{equation}
    \langle{x}\rangle_{N+1}=\langle{P}\rangle_{N+1}=1-\langle{P}\rangle_{N}
    =\frac{E_{R}}{1-\kappa} 
    \label{firstmoments}
\end{equation} 
and $\langle x\rangle_{N}=\langle v\rangle_{N}=\langle v\rangle_{N+1}=0$. 
For $n+m=2$, we find for the variances~\cite{armour1}
\begin{eqnarray}
    \delta x^{2}&=&\langle x^{2}\rangle-\langle x\rangle^{2}= 
    \frac{e V_{\rm ds}}{m \omega^{2}}\langle 
    P\rangle_{N}\langle P\rangle_{N+1}\label{xvariance}\\
    \delta v^{2}&=&\langle v^{2}\rangle=(1-\kappa)\frac{e V_{\rm ds}}{m}
    \langle P\rangle_{N}\langle P\rangle_{N+1} \label{vvariance}
\end{eqnarray}
and $\langle xv\rangle_{N}=\langle xv\rangle_{N+1}=0$, where the averaged 
probabilities are defined as
\[\langle P\rangle_{N(N+1)}=\int du\int dv\ P_{N(N+1)}(x,v,t)\]
and we have returned to the original, dimensionful coordinates. 

\begin{figure}[htbp]
\centering
\epsfig{file=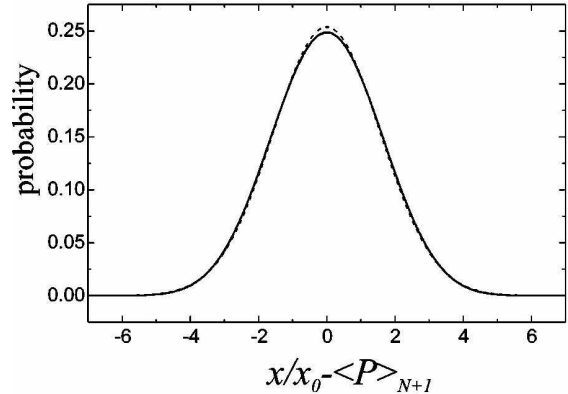,width=3.in}
\caption {Steady state probability distribution $P(x)$ for 
$\epsilon=0.3$ and $\kappa=0.1$~\cite{armour1}. 
The horizontal coordinates are in units of 
the shift $x_{0}$ and with origin at the steady state average position 
$x_{0}\langle P_{N+1} \rangle$. The numerical 
solution is given by the solid line and the Gaussian fit is 
the dashed line.}
\label{steadystategaussian}
\end{figure}

What can we learn from these solutions to the first few moments? For 
$\kappa<1$, the very existence of the solutions is  evidence 
that the SET damps the oscillator, bringing it to a steady state. For 
$\kappa>1$ on the other hand, the solutions are ill-defined (note 
that $\delta v^{2}$, $P_{N+1}<0$ if $\kappa>1$) 
suggesting that the coupled SET-oscillator system may be unstable in 
this regime. It is not possible to say anything more definite than 
this about the behaviour in the large $\kappa$ regime, however, since the 
approximations to the master equation made above (in particular the 
omission of the step functions and the tunneling rates 
to the right) break down when $\kappa$ is not small. Note from 
(\ref{xvariance}) and (\ref{vvariance}) 
that the position and velocity variances are related as
\begin{equation}
    \omega_{R}^{2}\delta x^{2}=\delta v^{2}=\frac{e V_{\rm ds}}{m}
    \langle P\rangle_{N}\langle P\rangle_{N+1},
    \label{equipartition}
\end{equation}
where the renormalized oscillator frequency is 
$\omega_{R}=\sqrt{1-\kappa}~\omega$.
Now recall that for a classical damped harmonic oscillator in contact 
with a thermal bath at temperature $T$, we have equipartition of energy:
$\frac{1}{2}m\delta v^{2}=\frac{1}{2}m\omega^{2}\delta 
x^{2}=\frac{1}{2}k_{\rm B}T$. Comparing with
equation~(\ref{equipartition}), this suggests that we can assign an effective 
temperature $T_{\rm SET}$ to the SET given by
\begin{equation}
    k_{\rm B}T_{\rm SET}=e V_{\rm ds} \langle P\rangle_{N}\langle 
    P\rangle_{N+1}.
    \label{SETtemperature}
\end{equation}
Further evidence that the SET behaves effectively as a 
thermal bath comes from examining the higher moments with $n+m=3, 4\ldots$. 
For $\kappa\ll 1$, it is found that the higher moments can be 
approximately decomposed into products of the lowest moments with $n+m=1, 2$, 
i.e.,
$\langle x\rangle$, 
$\langle x^{2}\rangle$, and $\langle v^{2}\rangle$. This means that the 
steady state probability distribution $P(x,v)~
[=P_{N}(x,v)+P_{N+1}(x,v)]$ for the oscillator is given by a 
Gaussian to a good approximation, as is the Maxwell-Boltzmann 
distribution describing a classical harmonic 
oscillator in contact with a thermal bath:
\begin{equation}
    P(x,v)=\frac{2\pi k_{\rm B}T}{m\omega} 
    \exp\left\{-\frac{m}{2 k_{\rm B}T}
    \left[\omega^{2} (x-\langle x\rangle)^{2}+v^{2}\right]\right\}.
    \label{gaussian}
\end{equation}
Figure~\ref{steadystategaussian} shows an example steady-state probability 
distribution for $P(x)~[=\int dv~P(x,v)]$ obtained by numerically 
solving the full master equations~(\ref{mastereq1}) and 
(\ref{mastereq2}) with parameter choices $\epsilon=0.3$ and 
$\kappa=0.1$. Also plotted is the Gaussian distribution obtained after 
integrating (\ref{gaussian}) over the velocity coordinate, with the 
steady state average position coordinate given by $\langle 
x\rangle=x_{0}\langle P_{N+1}\rangle$ [the dimensionful form of 
equation~(\ref{firstmoments})] and 
the temperature given by equation~(\ref{SETtemperature}). As can be 
clearly seen, the steady state distribution is closely approximated by the 
Gaussian distribution fixed using the analytically-derived effective 
temperature and steady state average position coordinate. 

\subsection{Mechanical resonator dynamics in the weak coupling regime} 
\label{weak}

In the previous section, we found that for weak coupling ($\kappa\ll 
1$) the SET appears to the oscillator in the steady state as a thermal 
bath with effective temperature 
$k_{\rm B}T_{\rm SET}=e V_{\rm ds} \langle P\rangle_{N}\langle P\rangle_{N+1}$.
Suppose now that the resonator is not in a steady state, e.g., it is 
given some initial displacement amplitude  and released, undergoing 
subsequent non-steady state motion. Does the SET still appear to the 
oscillator as a thermal bath? The following analysis addresses this 
question and as we shall learn, the SET indeed behaves as a bath, 
provided there is a wide separation in the oscillator 
and SET dynamics timescales ($\epsilon\ll 1$) in addition to weak 
coupling.

We seek some approximate way to solve the dimensionless master 
equations~(\ref{dimensionlessmastereq1}) and 
(\ref{dimensionlessmastereq2}) which takes advantage of the weak 
coupling condition $\kappa\ll 1$. The solution should furthermore be 
a good approximation for times much longer than the oscillator 
period $2\pi/\omega$ in order to establish that the SET damps the 
oscillator. The obvious procedure is to solve for the evolving 
probability distribution perturbatively using $\kappa$ as the small 
expansion parameter. Let us first put the master equations in the 
following concise, matrix form:
\begin{equation}
    \dot{\cal P}=\left({\cal H}_{0}+{\cal V}\right){\cal P},
    \label{schrodingereq}
\end{equation}
where ${\cal P}=\left(\begin{array}{c} P_{N}(x,v,t)\\P_{N+1}(x,v,t)\\ 
\end{array}\right)$,
\begin{equation}
    {\cal H}_{0}=
    \left(\begin{array}{cc}
    \epsilon^{2}x\frac{\partial}{\partial 
    v}-v\frac{\partial}{\partial x} & 0\\
    0 & \epsilon^{2}x\frac{\partial}{\partial 
    v}-v\frac{\partial}{\partial x}\\
    \end{array}\right)  +
    \left(\begin{array}{cc}
    -E_{R}-\kappa\langle x\rangle & E_{L}-\kappa\langle x\rangle\\
     E_{R}+\kappa\langle x\rangle & -E_{L}+\kappa\langle x\rangle\\
    \end{array}\right)
    \label{H0}
\end{equation}
and ${\cal V}={\cal V}_{1}+{\cal V}_{2}$, with
\begin{equation}
    {\cal V}_{1}=
    \kappa x\left(\begin{array}{cc}
    -1 & -1\\
    1 & 1\\
    \end{array}\right)
    \label{V1}
\end{equation}
and
\begin{equation}
    {\cal V}_{2}=
    \epsilon^{2} \frac{\partial}{\partial v} \left(\begin{array}{cc}
    \langle x\rangle & 0\\
     0 &\langle x\rangle -1\\
    \end{array}\right),
    \label{V2}
\end{equation}
where we have redefined the position coordinate such that its origin 
coincides with the steady-state value $\langle x\rangle\approx 
E_{R}$ [see equation~(\ref{firstmoments})]. 
Equation~(\ref{schrodingereq}) resembles the time-dependent 
Schr\"{o}dinger equation, but without the imaginary $i$ since it 
describes a classical and not a quantum system. The `Hamiltonian operator' 
${\cal H}_{0}$ gives the free, decoupled evolution of the independent 
oscillator and SET systems, while the operator 
${\cal V}={\cal V}_{1}+{\cal V}_{2}$ describes the 
interaction between the two systems with ${\cal V}_{1}$ giving the 
dependence of the SET tunneling rates on the oscillator position and ${\cal 
V}_{2}$ giving the SET island number dependence of the electrostatic force 
acting on the oscillator. 

Given the resemblance of equation~(\ref{schrodingereq}) to the 
Schr\"{o}dinger equation, we can consider applying approximation 
schemes that have been developed in quantum mechanics. We shall use a 
method developed for open quantum systems which is sometimes called 
the `self-consistent Born approximation' (SCBA). A system is open when it is 
coupled to another system with an infinite number of degrees of 
freedom. The latter, infinite system is commonly termed the 
`environment' or `reservoir' and the approximation method seeks to 
derive simpler, effective equations of motion for the finite system 
alone where the environment degrees of freedom have been integrated out. In our 
case, the finite system of interest is the mechanical resonator 
modeled as a harmonic oscillator, while the environment comprises the 
tunneling electrons in the SET. Applying the SCBA as described in 
section 3.1 of Ref.~\cite{paz1}, we obtain the following approximate effective 
equation of motion for the probability distribution $P_{\rm HO}(x,v,t)$ of the
oscillator:
\begin{eqnarray}
    &&\dot{P}(x,v,t)_{\rm HO}={\cal H}_{\rm HO} P_{\rm HO}(x,v,t)\cr
    &&+e^{{\cal H}_{\rm HO}t}{\rm Tr}_{\rm SET}\left[{\cal V}(t) {\cal P}_{\rm 
    SET}(0)\right]e^{-{\cal H}_{\rm HO}t}P_{\rm HO}(x,v,t)\cr
    &&-\int_{0}^{t}dt' ~e^{{\cal H}_{\rm HO}t} 
    {\rm Tr}_{\rm SET}\left[{\cal V}(t) {\cal P}_{\rm 
    SET}(0)\right]{\rm Tr}_{\rm SET}\left[{\cal V}(t') {\cal P}_{\rm 
    SET}(0)\right]e^{-{\cal H}_{\rm HO}t}P_{\rm HO}(x,v,t)\cr
    &&+\int_{0}^{t}dt' ~e^{{\cal H}_{\rm HO}t} 
    {\rm Tr}_{\rm SET}\left[{\cal V}(t){\cal V}(t') {\cal P}_{\rm 
    SET}(0)\right]e^{-{\cal H}_{\rm HO}t}P_{\rm HO}(x,v,t),
    \label{SCBA}
\end{eqnarray}
where ${\cal H}_{\rm HO}=\epsilon^{2}x\frac{\partial}{\partial v}-
v\frac{\partial}{\partial x}$ is the Hamiltonian operator for the free 
harmonic oscillator and 
${\cal V}(t)=e^{-{\cal H}_{0}t}{\cal V}e^{+{\cal H}_{0}t}$ 
is in the interaction picture. The initial, $t=0$  probability 
distribution is assumed to be a product state: ${\cal P}(0)=
P_{\rm HO}(x,v,0){\cal P}_{\rm SET}(0)$, where ${\cal 
P}_{\rm SET}(0)=\left(\begin{array}{c} P_{N}(0)\\P_{N+1}(0)\\ 
\end{array}\right)$. 

The influence of the SET on the
oscillator is contained in the `trace' terms, such as 
${\rm Tr}_{\rm SET}\left[{\cal V}(t) {\cal P}_{\rm SET}(0)\right]=
\sum_{\alpha=1}^{2}\sum_{\beta=1}^{2}{\cal V}_{\alpha\beta}(t) {\cal P}_{{\rm 
SET} \beta}(0)$. All such terms in (\ref{SCBA}) and their integrals 
with respect to $t'$ can evaluated analytically. The resulting 
explicit expression for (\ref{SCBA}) is rather complicated, involving
$t$-independent terms, $t$-dependent decaying terms of the form 
$e^{-t}$, as well as  decaying oscillatory terms of the form 
$e^{-t}\cos (\epsilon t)$ and $e^{-t}\sin (\epsilon t)$.  The terms 
also depend on the initial state ${\cal P}_{\rm SET}(0)$ of the SET.
However, if $\epsilon\ll 1$, then we can neglect all decaying terms 
involving $e^{-t}$ since, on timescales of order the mechanical 
period and longer,
$t\gtrsim\epsilon^{-1}$, such terms have a negligible effect on the 
 oscillator dynamics. Furthermore, the dependence on the initial 
state of the SET drops out. Neglecting the time-dependent decaying 
terms is usually called the `Markov' approximation, while the combined steps 
of the SCBA and Markov approximation are often simply called the `Born-Markov' 
approximation. As a result, we obtain the following much simpler
effective equation of motion for the oscillator: 
\begin{equation}
    \frac{\partial P}{\partial t}=
    \left[\epsilon^{2}x\frac{\partial}{\partial v}-
    v\frac{\partial}{\partial x} +\kappa\epsilon^{2}
    \frac{\partial}{\partial v}\left(v-x\right) +\epsilon^{4}\langle 
    P\rangle_{N}\langle P\rangle_{N+1}\frac{\partial}{\partial v}
    \left(\frac{\partial}{\partial 
    v}+\frac{\partial}{\partial  x}\right)\right] P,
    \label{fokkerplanck}
\end{equation}
where $\langle P\rangle_{N}$ and 
$\langle P\rangle_{N+1}$ ($=1-\langle P\rangle_{N}\approx E_{R}$) are 
the steady state SET island electron number 
probabilities [see equation~(\ref{firstmoments})] and
we have dropped the `HO' subscript on the oscillator probability 
distribution $P(x,v,t)$ for notational convenience. 
Expressing~(\ref{fokkerplanck}) in terms of dimensionful coordinates, 
we obtain
\begin{equation}
     \frac{\partial P}{\partial 
     t}=\left[\omega_{R}^{2}x\frac{\partial}{\partial v} 
     -v\frac{\partial}{\partial x}+\gamma_{\rm SET}\frac{\partial}{\partial v}v
     +\frac{\gamma_{\rm SET} k_{\rm B}T_{\rm SET}}{m} 
     \frac{\partial^{2}}{\partial v^{2}}\right] P,
     \label{fokkerplanck2}
\end{equation}
where recall the renormalized oscillator frequency is 
$\omega_{R}=\sqrt{1-\kappa}~\omega$, the damping rate is
\begin{equation}
    \gamma_{\rm SET}=\kappa\epsilon\omega
    \label{dampingrate}
\end{equation}
and the effective SET temperature $T_{\rm SET}$ is defined in 
(\ref{SETtemperature}).

Equation~(\ref{fokkerplanck2}), which goes by the name `Klein-Kramers' 
or `Fokker-Planck' equation~\cite{risken}, describes the brownian motion of a 
harmonic oscillator interacting with a thermal bath; the oscillator experiences 
a damping force due to the SET [the third term on the right-hand-side of 
equation~(\ref{fokkerplanck2})]  
with quality factor $Q_{\rm SET}=\omega/\gamma_{\rm SET}=1/(\kappa\epsilon)$ 
(since $\kappa\ll 1$, the renormalized frequency can be 
replaced by the unrenormalized frequency in the definition of the 
quality factor) and 
an accompanying Gaussian distributed thermal fluctuating force [the fourth term 
on the right-hand-side of (\ref{fokkerplanck2}), called the `diffusion' 
term]. The $\partial^{2} P/\partial v\partial x$ term in 
equation~(\ref{fokkerplanck}) (called the `anomalous diffusion' term 
in Ref.~\cite{paz1})  has been neglected because it is of order 
$\epsilon$ smaller than the diffusion term when time is expressed in 
units of the oscillator period; it should have only a small effect on 
timescales of order the mechanical period or longer. The first 
moment of the fluctuating force which gives rise to the diffusion 
term vanishes, $\langle F(t)\rangle =0$, while its second moment is
\begin{equation}
    \langle F(t) F(t')\rangle =
    \frac{2\gamma_{\rm SET} k_{\rm B}T_{\rm SET}}{m}\delta(t-t').
    \label{correlation}
\end{equation}
The delta function in equation~(\ref{correlation}) signifies that  
the kicks inflicted on the oscillator by the SET are uncorrelated, no 
matter how small is the distinct time interval separating the kicks. 
This amounts to taking the limit $\tau_{\rm tunnel}\rightarrow 0$, 
which is justified provided $\tau_{\rm tunnel}\ll 2\pi/\omega$ 
(equivalently $\epsilon\ll 1$) and we are 
interested only in the oscillator dynamics on timescales $t\gtrsim 
2\pi/\omega$ as discussed above. The condition $\epsilon\ll 1$ appears 
necessary for the oscillator to perceive the SET as a thermal bath; 
if the condition did not hold, then it would be possible to 
infer from the oscillator dynamics that it was coupled to a tunneling 
electron device and not to a many degree-of-freedom system in a state 
of thermal equilibrium. 

The second and third terms on the right-hand-side of equation~(\ref{SCBA}) 
involving traces of single 
interaction potential operators ${\cal V}(t)$ vanish by virtue of the 
Markov approximation and also the fact that the position coordinate 
was originally redefined such that the origin coincides with the 
steady-state value. The damping and diffusion terms in 
equation~(\ref{fokkerplanck}) 
arise from the fourth term on the right-hand-side of 
equation~(\ref{SCBA}) involving the trace of the 
interaction potential operator product  ${\cal V}(t){\cal V}(t')$. 
More specifically, only products of the form ${\cal V}_{2}(t){\cal 
V}_{1}(t')$ and ${\cal V}_{2}(t){\cal 
V}_{2}(t')$ [see definitions~(\ref{V1}) and (\ref{V2})] 
give non-zero contributions, with the damping and frequency 
renormalization terms arising from the 
first product and the diffusion and anomalous diffusion terms arising from the 
second product. 
Thus, the effective thermal bath 
description for the SET crucially requires one to take into account 
not only the influence of the SET on the oscillator (${\cal V}_{2}$), but 
also the influence of the oscillator on the SET as well (${\cal 
V}_{1}$). 

In our analysis of the coupled SET-oscillator dynamics, we 
have neglected the coupling between the  oscillator and its 
external thermal bath. Incorporating the external bath into the 
equations of motion is straightforward: we just add the term 
\begin{equation}
    \gamma_{\rm ext}\frac{\partial (vP)}{\partial v}+
    \frac{\gamma_{\rm ext}k_{\rm B}T_{\rm 
    ext}}{m}\frac{\partial^{2}P}{\partial v^{2}}
    \label{externalKramers}
\end{equation}
to the above Fokker-Planck equation~(\ref{fokkerplanck2}), where we 
assume again for simplicity that the influence of the external bath on the 
oscillator is uncorrelated and memoryless. The 
damping and fluctuating force terms due to the SET and external bath 
can be combined respectively to yield a single damping term with effective quality 
factor $Q_{\rm eff}^{-1}=Q_{\rm SET}^{-1}+Q_{\rm ext}^{-1}$ and a 
single fluctuating force term with effective temperature
\begin{equation}
    T_{\rm eff}=Q_{\rm eff} \left(\frac{T_{\rm SET}}{Q_{\rm SET}}+
    \frac{T_{\rm ext}}{Q_{\rm ext}}\right).
    \label{effectivetemp}
\end{equation}

\begin{figure}[htbp]
\centering
\epsfig{file=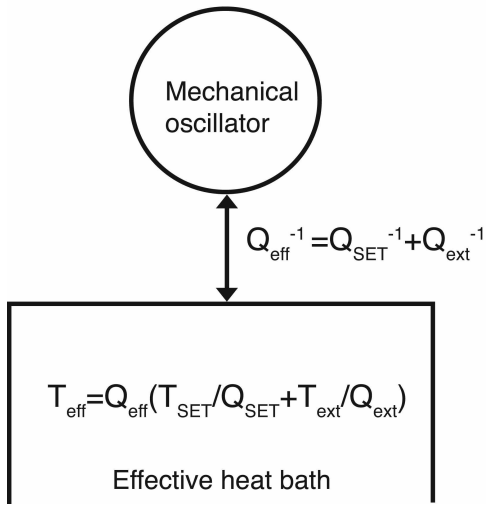,width=2.5in} 
\caption {Scheme of the SET-oscillator system under the conditions 
of weak coupling and wide separation of dynamic timescales between the 
SET and oscillator. The oscillator undergoes 
thermal brownian motion, behaving as if in contact with a thermal bath at 
temperature $T_{\rm eff}=Q_{\rm eff} \left({T_{\rm SET}}/{Q_{\rm SET}}+
{T_{\rm ext}}/{Q_{\rm ext}}\right)$, where 
$Q_{\rm eff}^{-1}=Q_{\rm SET}^{-1}+Q_{\rm ext}^{-1}$.}
\label{weakcouplingnems}
\end{figure}

Summarizing so far, we have learned that provided the mechanical 
oscillator and SET 
are weakly coupled ($\kappa\ll 1$) and, furthermore, provided the SET dynamics 
occurs on much shorter timescales than the oscillator dynamics 
($\epsilon\ll 1$), then the oscillator behaves effectively as if it is 
in contact with a thermal bath, with figure~\ref{genericnems} being 
replaced by the much simpler figure~\ref{weakcouplingnems}. 

For typical SET device parameters and micron-scale resonators with 
frequencies of the order of a few megahertz, separated by a resonator-island 
electrode gap of about $0.1~\mu{\rm m}$, one obtains effective 
SET temperatures of around $1{\rm K}$ and effective SET quality factors 
ranging from around $10^{4}$ down to $10^{2}$ as the gate voltage 
increases from $1~{\rm V}$ up to $10~{\rm V}$~\cite{armour1}. External 
temperatures and quality factors for micron-scale resonators
are of the order $100~{\rm mK}$ and $10^{4}$--$10^{5}$, respectively,
so that it should be 
possible to measure the effects of the SET temperature and quality 
factor on the mechanical resonator.
Experiments are underway in Keith Schwab's group to probe 
the coupled dynamics of the SET and mechanical resonator~\cite{schwab1}. 
They have seen evidence in several devices that the SET behaves as an 
effective bath, heating the resonator and causing damping which 
increases with gate voltage. This is the first time  
that the back action of electronic shot noise on a nanomechanical 
resonator has been observed.  Note, 
however, that they work with superconducting SETs (SSETs), where tunneling 
processes involve  Cooper pairs as well, so that the 
above formulae for $T_{\rm SET}$ and $Q_{\rm SET}$ do not apply. It is 
important then to also analyze the SSET-mechanical resonator coupled 
dynamics~\cite{clerk3,blencowe4}. 

\section{Effective equilibrium dynamics of other NEMS} \label{NEMS}

\begin{figure}[htbp]
\centering
\epsfig{file=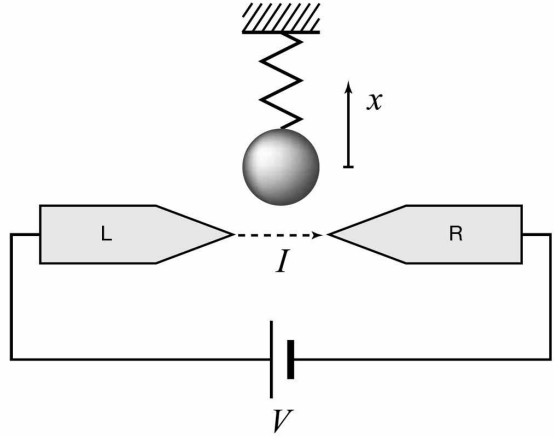,width=3in} 
\caption {Oscillator coupled to an electrical tunnel 
junction~\cite{mozyrsky1}. As a 
result of an applied voltage $V$, electrons will tunnel from the 
right (R) to 
left (L) electrode, with tunnel rates depending on the position $x$ of a 
nearby mechanical oscillator. The electromechanical coupling can be 
achieved by putting a net non-zero charge (or voltage) on a metallized 
mechanical oscillator.}
\label{mozyrsky}
\end{figure}

In this section, we consider the effective equilibrium 
dynamics of some of the other NEM devices described in 
section~\ref{survey}. One of the first theoretical investigations 
to establish that a far-from-equilibrium electronic device  can behave
like an effective thermal bath was conducted by Dima Mozyrsky and Ivar 
Martin at Los Alamos~\cite{mozyrsky1}. They considered a model device 
comprising a single, electrical tunnel junction  with tunneling rates 
depending on the position coordinate of a nearby mechanical 
oscillator, as illustrated in figure~\ref{mozyrsky}.
Solving the quantum Schr\"{o}dinger equation for the coupled, 
tunneling electrons-oscillator system, they found under the conditions 
of weak coupling and Markovian approximation that the oscillator 
behaves effectively as if it is in contact with a thermal bath, 
with the tunnel junction characterized by a damping rate and a temperature
given by
\begin{equation}
    k_{\rm B} T_{\rm TJ}=eV/2,
    \label{TempTJ}
\end{equation}
where $V$ is the voltage applied across the tunnel electrodes.

Thus, despite the differences between the tunnel junction and 
SET (two tunnel junctions in series with gated central island), 
both can behave effectively as 
a heat bath. In fact, the 
similarity in their statistical properties goes further. It turns out 
that their respective temperature expressions (\ref{SETtemperature}) and 
(\ref{TempTJ}) can be commonly equated to the ensemble-averaged energy 
gained by an electron due to tunneling across a junction~\cite{armour1}. 
For the tunnel junction, the averaged energy gained is $eV/2$, assuming zero 
electron temperature and constant density of states in the 
electrodes, as well as energy-independent tunneling 
matrix~\cite{mozyrsky1,smirnov1}. For the SET, with the same asumptions
the average energy 
gained by a tunneling electron is $(E_{R}/2) \langle P\rangle_{N}+   
(E_{L}/2) \langle P\rangle_{N+1}\approx eV_{\rm ds}\langle P\rangle_{N}
\langle P\rangle_{N+1}$ for $\kappa\ll 1$, where we have used equations
(\ref{dimensionlessenergy1}), (\ref{dimensionlessenergy2}) and 
(\ref{firstmoments}).

While  the mechanically compliant 
tunnel electrode illustrated in figure~\ref{schwabe} closely 
resembles the just-considered model tunnel junction 
(figure~\ref{mozyrsky}), there is a 
difference in that the tunneling electrons will give a momentum 
recoil to the mechanically compliant electrode, which is of course 
absent for the fixed electrode. To our knowledge, there has yet to be 
a comprehensive analysis of the coupled dynamics of tunneling 
electrons and mechanically compliant electrode, including the effects 
of momentum recoil. Nevertheless, one might expect that under the 
appropriate conditions of weak coupling and wide separation of 
dynamical timescales between the mechanical and electronic degrees of 
freedom, the effective thermal bath description again arises. 

As mentioned in section~\ref{survey}, the electronic and mechanical 
dynamics are strongly coupled in the classical charge shuttle 
(figure~\ref{scheible}); in order for a current to flow, the 
nanomechanical island electrode must shuttle charge between the drain 
and source electrodes, so that the dynamical timescales of the mechanical and 
electronic degrees of freedom are comparable. Because it is not 
possible to have a wide separation of timescales, it is unlikely that 
there is a regime in which the electron tunneling dynamics is 
effectively that of a thermal bath, as perceived by the mechanical 
shuttle. The only possibility of separating the timescales is to make 
the drain-source electrode gap smaller than the electron tunneling 
length (so that a tunneling current can flow without the nanomechanical island 
electrode having to move from drain to source electrode), likely an 
unrealisable regime for nanoscale mechanical resonators that must fit in the 
gap.

Thus, provided the electron dynamics is stochastic, the
electromechanical coupling strength is weak and the mechanical 
dynamics is much slower than the electron dynamics, the above examples 
suggest that the electronic device can be effectively replaced by a 
thermal bath for the reduced dynamics of the coupled mechanical 
resonator, so that figure~\ref{genericnems} can be replaced by 
figure~\ref{weakcouplingnems} (but now refering to a generic NEMS). How might one go 
about providing a  proof of 
this conjecture for a general class of electromechanical systems?
A possible direction is 
suggested by the Born-Markov approximation method applied to the 
SET-resonator system in the previous section~\ref{weak}. In fact, the 
derivation of the Fokker-Planck equation (that ensures the effective 
thermal bath description) does not depend on the detailed form of the 
matrices appearing in the definitions (\ref{V1}) and (\ref{V2})
of the interaction operators ${\cal V}_{1}$  and ${\cal V}_{2}$, 
respectively; 
the derivation of the Fokker-Planck equation should carry through for 
additional and different types of tunneling process, e.g., for a 
superconducting SET~\cite{blencowe4}. Of course, the detailed 
expressions for the renormalized frequency, damping constant and 
temperature will differ from one device to another. The important 
necessary steps that lead to the Fokker-Planck equation are the second order 
expansion in the interaction operator ${\cal V}$ 
(Born approximation--requires weak coupling) and the neglect of time-dependent 
decaying terms in the electron dynamics (Markov 
approximation--requires fast electron dynamics as compared with 
mechanical oscillator period). A suitable starting point for a general 
proof would be to write down a Boltzmann/master equation involving a 
probability density function $P_{\{k\}}(x,v,t)$ where $(x,v)$ are the 
oscillator's classical position/velocity coordinates and ${\{k\}}$ 
denotes some appropriate choice of electronic coordinates. The general 
interaction operator ${\cal V}$ should involve both position and 
velocity (to account for momentum recoil) dependent terms.  The proof 
would then involve applying the Born-Markov approximation to this 
master equation, recovering the Fokker-Planck equation. An important 
proviso, however, is that there is no guarantee that the electronic 
device can necessarily damp the oscillator. The tunneling dynamics 
may be such that it is more likely for an electron to give rather 
than to take energy from an oscillator, resulting in unstable coupled 
dynamics. This instability manifests itself through `negative damping' 
and `negative 
temperature' coefficients in the Fokker-Planck equation. Such an 
instability occurs in the case of the superconducting SET for certain 
ranges of gate and drain-source voltage~\cite{clerk3,blencowe4}. 

A proof of the effective thermal bath description
along the above lines has recently been demonstrated by Aashish 
Clerk at McGill University~\cite{clerk2}. He gives a quantum 
treatment, with the electronic device modeled as a generic linear 
response detector of the quantum oscillator's position (essentially 
the quantum version of figure~\ref{genericnems}), to which it is 
weakly coupled. The proof is in part a generalization of Mozyrsky and Martin's 
analysis of the tunnel junction~\cite{mozyrsky1}. 
However, as described above, when the coupled dynamics under 
consideration is classical it should be possible to 
give a similar proof  within the framework of classical master equations. Indeed, in Mozyrsky 
and Martin's tunnel junction analysis, the condition $eV\gg\hbar\omega$ 
is assumed, which from equation (\ref{TempTJ}) amounts to taking the classical 
limit $k_{\rm B}T_{\rm TJ}\gg \hbar\omega$ for the oscillator. Certainly, a 
classical derivation 
of the effective thermal bath description for NEMS avoids having to 
simultaneously deal with the complicating (but fascinating) issue 
of how classical dynamics arises by approximation from quantum 
dynamics.    

\section{Conclusions} \label{conclusion}

We have presented a brief overview of the field of NEMS research, with 
an emphasis on the classical effective dynamics of a nanomechanical 
resonator due to its interaction with a tunneling electron system. Under the 
conditions of weak coupling and the characteristic mechanical dynamics 
occuring on much longer timescales than the tunneling electron 
dynamics, the electron system can behave effectively as a heat 
bath, causing the resonator to undergo thermal brownian motion. 

As they have been stated, the conditions for the resonator to perceive the 
electronic system as a thermal bath are rather general. It 
is natural therefore to wonder whether there are other 
(semi)classical examples  
involving (not necessarily nanoscale) mechanical 
oscillators 
coupled to (not necessarily electronic) 
far from equilibrium 
systems with many degrees of freedom, such that the oscillator undergoes 
effective thermal brownian motion. 
Many such examples have in fact been considered. One 
example is laser Doppler cooling of a harmonically trapped 
ion~\cite{phillips1,foot1}. The laser 
is tuned below the atomic resonance, so that the Doppler shift makes 
it more likely for the atom to absorb a photon with momentum 
oppositely directed to that of the atom than in the same direction, 
resulting in damping of the atom's motion. Despite the fact that the 
laser radiation field is obviously not a thermal equilibrium state, 
nevertheless under certain weak coupling and wide 
timescale separation conditions, the atom's dynamics is 
described by a Fokker-Planck equation~\cite{javanainen1}, 
with effective temperature much lower than the ambient temperature: 
the laser cools the trapped ion. Another, recently considered 
macroscopic oscillator example~\cite{ojha1} 
involves placing a sphere (ping-pong ball) in an upward gas-flow. 
It was found that the 
sphere behaves effectively as a harmonic oscillator undergoing thermal 
brownian motion, despite the far-from-equilibrium, turbulent state of the  
gas molecules. A derivation of this behavior might 
start with a Boltzmann equation for a single 
massive particle undergoing collisions with a dilute gas of small mass 
particles, and then identifying the relevant small dimensionless coupling 
and timescale parameters for a subsequent series expansion of the 
Boltzmann equation. However,
the recovery of a Fokker-Planck equation 
is not straightforward, owing to the 
relevance of the non-trivial turbulent gas dynamics for the effective 
dynamics of the oscillator. Yet another, related macroscopic example 
involves immersing a torsion oscillator in a far-from-equilibrium,
vibration-fluidized 
granular medium~\cite{d'anna1,umbanhowar1,baldassarri1}. 
Again, the granular medium was found to behave 
effectively as a thermal bath, with the oscillator undergoing 
thermal brownian motion. 

Returning to the subject of NEMS, there are many aspects of their 
dynamical behaviour still to be understood. In the strong coupling 
regime (e.g., $\kappa\gtrsim 1$ in the case of the SET-oscillator system), 
little is known in general about the effective dynamics of the oscillator
and also how the coupled dynamics manifests itself in the effective behavior 
of the electronic system (e.g., current, current noise, etc.). 
Furthermore, little is known in general 
about the quantum dynamics of NEMS and how (semi)classical dynamics 
arises as a result of dephasing within the electronic system, as well 
as due to the external environment of the nanomechanical resonator.

\section*{Acknowledgments}
We especially thank A. Armour, A. Clerk, I. Martin, R. Onofrio and K. Schwab 
for very helpful discussions. We also thank M. Plenio for providing 
the opportunity to write this review.

\section*{Biography}

{\it Miles  Blencowe received his B.Sc. (1986) and 
Ph.D. (1989) degrees in theoretical physics at Imperial College. 
He subsequently held postdoctoral fellowships at the University of 
Cambridge (1989-1991), the University of Chicago (1991-1993) and the 
University of British Columbia (1993-1994). 
Following a senior research associate position at Imperial College 
(1994-1999), he moved to Dartmouth College where he is currently an 
associate professor in the Department of Physics and Astronomy. His 
research interests lie in the areas of mesoscopic physics, 
quantum information theory applied to solid state systems, and 
non-equilibrium statistical mechanics.}

\end{document}